\documentclass[12pt,preprint]{aastex}

\begin{document}
\title{The Creation of Haumea's Collisional Family}
\shortauthors{Schlichting & Sari}
\shorttitle{Haumea's Collisional Family}
\author{Hilke E. Schlichting\altaffilmark{1} and Re'em Sari\altaffilmark{1,2} }
\altaffiltext{1}{California Institute of Technology, MC 130-33, Pasadena, CA
  91125} 
\altaffiltext{2}{Racah Institute of Physics, Hebrew University,
  Jerusalem 91904, Israel}
\email{hes@astro.caltech.edu, sari@tapir.caltech.edu}

\begin{abstract} 
Recently, the first collisional family was discovered in the Kuiper belt. The
parent body of this family, Haumea, is one of the largest objects in the
Kuiper belt and is orbited by two satellites. It has been proposed that the
Haumea family was created from dispersed fragments that resulted from a giant
impact. This proposed origin of the Haumea family is however in conflict with
the observed velocity dispersion between the family members ($\sim 140~$m/s)
which is significantly less than the escape velocity from Haumea's surface
($\sim 900~$m/s). In this paper we propose a different formation scenario for
Haumea's collisional family. In our scenario the family members are ejected
while in orbit around Haumea. This scenario, therfore, gives naturally rise to
a lower velocity dispersion among the family members than expected from direct
ejection from Haumea's surface. In our scenario Haumea's giant impact forms a
single moon that tidally evolves outward until it suffers a destructive
collision from which the family is created. We show that this formation
scenario yields a velocity dispersion of $\sim 190 \rm{m/s}$ among the family
members which is in good agreement with the observations. We discuss an
alternative scenario that consists of the formation and tidal evolution of
several satellites that are ejected by collisions with unbound Kuiper belt
objects. However, the formation of the Haumea family in this latter way is
difficult to reconcile with the large abundance of Kuiper belt binaries. We
therefore favor forming the family by a destructive collision of a single moon
of Haumea. The probability for Haumea's initial giant impact in todays Kuiper
belt is less than $ 10^{-3}$. In our scenario, however, Haumea's giant impact
can occur before the excitation of the Kuiper belt and the ejection of the
family members afterwards. This has the advantage that one can preserve the
dynamical coherence of the family and explain Haumea's original giant impact,
which is several orders of magnitude more likely to have occurred in the
primordial dynamically cold Kuiper belt compared to the dynamically excited
Kuiper belt today.
\end{abstract}

\keywords{comets: general---Kuiper Belt---minor planets, asteroids --- solar
system: formation}

\section{INTRODUCTION} 
Collisions are thought to have played a major role in the Kuiper belt ever
since its formation \citep[e.g.][]{DF97,SC97,KL99,GLS02,PS05}. This idea is
supported further by the recent discovery of the first collisional family in
the Kuiper belt \citep{BB07}. Haumea (also known as $\rm{2003~EL_{61}}$), one
of the largest Kuiper belt objects (KBOs), is thought to have undergone a
giant impact that gave rise to Haumea's rapid rotation with a spin period of
only 4 hours \citep{RB06} and that created its multiple satellite system
\citep{BB05,BvD06} and collisional family \citep{BB07}. The family of KBOs
(1995 SM$_{\rm{55}}$, 1996 TO$_{\rm{66}}$, 1999 OY$_{\rm{3}}$, 2002
TX$_{\rm{300}}$, 2003 OP$_{\rm{32}}$, 2003 UZ$_{\rm{117}}$, 2005
CB$_{\rm{79}}$, 2005 RR$_{\rm{43}}$) was linked to Haumea because its members
display surface properties and orbits similar to those of Haumea. It has been
proposed that this family of KBOs are collisional fragments of the ejected ice
mantel of Haumea that were produced and ejected in Haumea's giant impact
\citep{BB07}. However, the velocity dispersion between the family members is
only $\sim 140$m/s which is unusually small for fragments of disruptive
impacts which should typically be ejected with a velocity comparable to the
escape velocity (i.e. $\sim 900$m/s for Haumea) \citep{BA99,N06}. In addition,
simulations suggest that high velocity giant impacts lead to either the
formation of disk of satellites or the dispersion of collisional
fragments. The simultaneous creation of multiple satellites and the dispersion
of collisional fragments has not been seen \citep{MR97,BA99,AA04,C04,C05};
one should bear in mind, however, that none of the simulations tried to
specifically explain Haumea's giant impact.

In this paper we propose and discuss a different formation scenario for the
origin of Haumea's collisional family. In our scenario the family members are
ejected while in orbit around Haumea. Ejecting the family members while in
orbit around Haumea has the advantage that it naturally gives rise to a lower
velocity dispersion among the family members than a direct ejection of
fragments from Haumea's surface and, in addition, it aids in explaining
Haumea's initial giant impact.

This paper is structured as follows.  In \S \ref{5s1} we introduce our
definitions and assumptions. We give a detailed account of our model for the
formation of Haumea's collisional family in \S \ref{5s2}. \S \ref{5s3} is
concerned with Haumea's giant impact.  Discussion and conclusions follow in \S
\ref{5s4}.

\section{DEFINITIONS AND ASSUMPTIONS}\label{5s1}
The Haumea family currently consists of Haumea and eight additional KBOs. The
family members have typically an eccentricity of $\sim 0.12$ and an
inclination of $\sim 27^{\circ}$. The actual masses of the family members are
uncertain since Haumea is the only object in its family with a measured
albedo. Haumea's visible albedo is with about $70\%$ \citep{RB06,SG108} among
the highest in the solar system. In addition, lower limits for the visible
geometric albedo of family members $\rm{2002~TX_{300}}$, $\rm{1995~SM_{55}}$
and $\rm{1996~TO_{66}}$ were determined to be $19\%$ \citep{OSM04,GN05},
$6.7\%$ and $3.3\%$ \citep{AB04,GN05} respectively. Given the common origin of
the Haumea family and their similar surface characteristics with strong water
ice absorption features it is likely that the family members have, like
Haumea, high albedo surfaces \citep{RS08}. We will therefore
assume that all family members have an albedo similar to that of Haumea and we
calculate the masses of the family members from absolute magnitudes from
\citet{RB07} and references within.

Estimates from current Kuiper Belt surveys yield for the mass surface density
$\Sigma \sim 3 \times 10^{-4} \rm{g~cm^{-2}}$ for KBOs of $\sim
100~\rm{km}$ in size \citep{PK08,FH108,FK108,TB03,TJL01}. We use this value of
$\Sigma$, assuming that no 100km sized objects were lost from the Kuiper belt
after it was dynamically excited. We use a power law distribution $N(r)
\propto r^{1-q}$ with power-law index $q \sim 4$ \citep{BTA04,FH08,FK09} when
estimating the number density of objects smaller than $\sim 100$km in this
paper.

For simplicity, we define symbols and their numerical values that will be used
throughout this paper in Table \ref{5t1}.
\begin{deluxetable}{cccccc}\label{5t1}
\tabletypesize{\scriptsize} \tablecaption{Definition of Symbols}
\tablewidth{0pt} \tablehead{ \colhead{Symbol} & \colhead{Value} &
\colhead{Definition} } \startdata M& $4.2 \times 10^{24}$g& mass of Haumea
\citep{BB05}\\ R& 694km& mean radius of Haumea\tablenotemark{a}\\ $\Omega_B$&
$9.2 \times 10^{-4}$rad/s& angular break up velocity of Haumea\\ $v_B$&
635m/s& break up velocity of Haumea\\ $\Omega$& $7.1 \times 10^{-10}$rad/s&
angular velocity around the sun\tablenotemark{b}\\ $v_{disp}$& 3km/s& velocity
dispersion in the scattered Kuiper belt\\ $\Sigma$& $3 \times
10^{-4}\rm{g/cm^2}$& Kuiper belt mass surface density of $\sim
100$km sized bodies\\ $m_s$& $\sim 2 \times 10^{20} - 3 \times 10^{22}$g& mass range
of Haumea's family member\tablenotemark{c}\\ 
\enddata
\tablenotetext{a}{for a density of $3\rm{g/cm^3}$ \citep{RB06}}
\tablenotetext{b}{evaluated at 43AU} \tablenotetext{c}{derived from magnitude
  difference between Haumea and the family members \& assuming same albedo as
  Haumea and a density of $1\rm{g/cm^3}$, magnitudes are taken from \citet{RB07} and references within}
\end{deluxetable}

\section{THE FORMATION OF HAUMEA'S COLLISIONAL FAMILY}\label{5s2}

\subsection{Formation of a Single Satellite and Ejection by destructive Satellite Collision}
\begin{figure}
\plotone{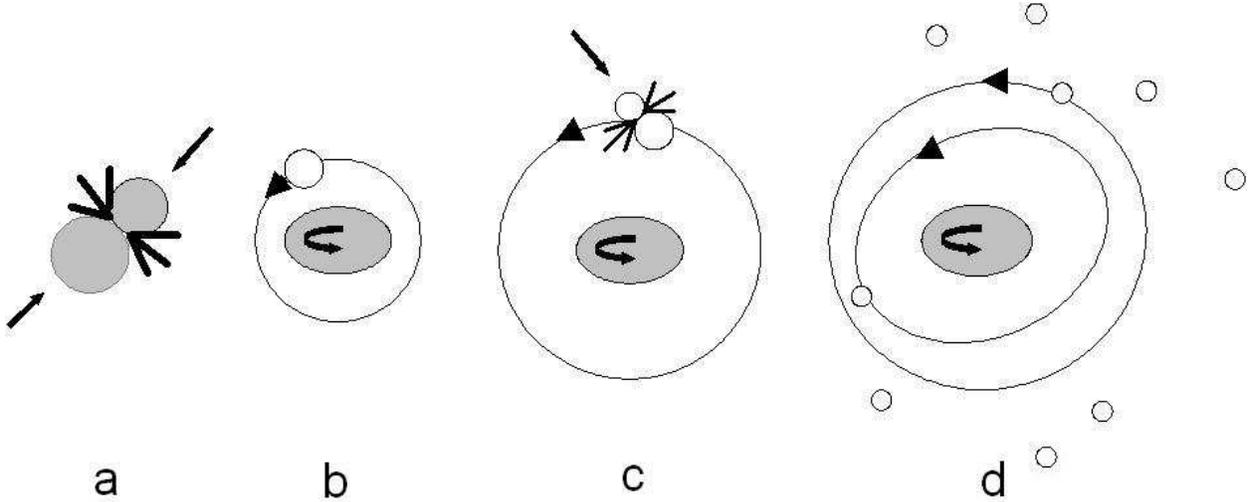} 
\caption{Cartoon of our model for the formation of Haumea's collisional
  family. From left to right; first, Haumea suffers a giant impact (a). This
  collision gives rise to Haumea's fast, $4$ hour spin period and ejects
  material that accumulates into a tightly bound satellite around Haumea
  (b). The newly formed satellite undergoes tidal evolution that increases its
  orbital separation from Haumea. Haumea's satellite suffers a destructive
  collision with an unbound KBO (c). This collision creates and ejects the
  family and forms the two moons (d).  }
\label{5fig1}
\end{figure}

Our formation scenario for Haumea's collisional family can be divided into
three steps. First, Haumea suffers a large collision. This collision gives
rise to Haumea's fast, $4$ hour spin period and ejects material that
accumulates into a tightly bound satellite around Haumea. Second, the newly
formed satellite undergoes tidal evolution that increases its orbital
separation from Haumea. Third, the satellite suffers a destructive collision
with an unbound KBO which creates and ejects the collisional family
(see Fig.\ref{5fig1}). In this case, the typical velocity dispersion
of the family will be of the order of the escape velocity from the satellite
which is $\sim 190m/s$ as will be shown below.

Starting with a tightly bound satellites around Haumea, the tidal
evolution timescale for a satellite of mass, $m_s$, to evolve from an initial
separation of a few times Haumea's radius, $R$, to a separation $a \gg R$ is
given by
\begin{equation}\label{5e31}
\tau_{tidal}=\left( \frac{2}{39}\right) \left(\frac{Q}{k}\right)
\left(\frac{M}{m_s}\right) \left(\frac{v_{B}}{v(a)}\right)^{13}\Omega_B^{-1}
\end{equation}
where $v(a)$ is the orbital velocity of the satellite with semi-major axis,
 $a$. $Q$ is the tidal dissipation function and $k$ the tidal Love number of
 Haumea. We refer the reader to table 1 for the definitions of the
 remaining symbols.

The satellite suffers a destructive collision with unbound KBOs at a rate
\begin{equation}\label{5e32}
\mathcal{R}_{coll}\sim \frac{\Sigma\Omega}{\rho r^3} r_s^2
\end{equation}
where $r$ is the radius of the `bullet' that can break up the satellite by
collisions. The initial satellite needs to have been at least as large as all
the identified family members combined (including Hi'iaka and Namaka but with
the exclusion of Haumea) which yields a satellite radius of $\sim
260$km. Bodies of this size are predominantly held together by their own
gravity. We can estimate the bullet size needed for satellite-break-up by
considering energy and momentum conservation of the shock that propagates from
the impact point to the antipode of the target. Since energy and momentum
conservation represent two limiting cases for impact processes \citep{H94}; we
estimate the range of bullet sizes needed for satellite-break-up to be
\begin{equation}\label{5e33}
\left(\frac{G \rho r_s^5}{v_{disp}^2}\right)^{1/3} < r < \left(\frac{G
  \rho r_s^8}{v_{disp}^2}\right)^{1/6}
\end{equation}
where the lower limit is derived by requiring that the kinetic energy of the
bullet be equal to the total gravitational energy of the target and the upper
limit results from conservation of momentum. We refer the reader to
\citet{PS05} for a detailed derivation and discussion of these two destruction
criteria for strengthless bodies. Evaluating equation (\ref{5e33}) yields
bullet sizes of $20 \rm{km} < r < 70 \rm{km}$. Substituting expression
(\ref{5e33}) into equation (\ref{5e32}) we find that the typical timescale for
Haumea's satellite to suffer a destructive collision is $ 2 \rm{Gyr} <
\tau_{coll} < 80 \rm{Gyr}$. The timescale for satellite-break-up by a
collision and the consequential formation of the family ranges therefore from
a few to tens of Gyrs. The actual $\tau_{coll}$ is most likely closer to the
tens of Gyrs since the majority of the binaries in the Kuiper belt should have
been destroyed by collisions otherwise. This, therefore, implies a probability
for a Haumea-type family forming event of $\sim 10 \%$ over the age of the
solar system. Estimates of the current Kuiper belt population indicate that
there should be about 10 objects with radii as larger as Haumea
\citep{TJL01,BTA04,FH08,FK09}. The abundance of Haumea-sized objects and the
ubiquity of collisionally formed satellites around the largest KBOs
\citep{BDB06,BS07} makes our formation scenario consistent with having one
collisional family for Haumea-sized objects in the Kuiper belt. However, we
expect additional collisional families that are associated with smaller parent
bodies. For example, we estimate that there are about 300 bodies in the Kuiper
Belt with sizes similar to that of the initial Haumea satellite. Given our
estimated satellite destruction probability of about 10\% we expect, to an
order of magnitude, about $\sim$ 30 collisional families that originated from
$\sim$ 260~km parent bodies. These families, however, would be harder to find
without a larger Haumea-type object and might lack any unique spectral
signatures that led to the identification of the Haumea family. The typical
velocity dispersion between the family members that are produced in the
satellite-break-up discussed above is of the order of the escape velocity of
the initial satellite. For a satellite radius of $\sim 260 \rm{km}$ and a
density of $1 \rm{g/cm^3}$ we have an escape velocity from the satellite,
$v_{esc}$, of $ \sim 190 \rm{m/s}$. Simulations of disruptive impacts on ice
and basalt targets find typical ejection velocities of $\sim 0.7 v_{esc}$ for
the largest remnant and impact velocities of 3km/s \citep{BA99}. It could
therefore be that the actual velocity dispersion of the family from a
disruptive impact is somewhat smaller than the $\sim 190$m/s estimated here.

Finally one needs to compare the escape velocity from the satellite,
$v_{esc}$, with its orbital velocity around Haumea, $v(a)$. A
satellite-break-up only leads to ejection from the Haumea system, and
therefore to the formation of a collisional family, if the tidal evolution has
increased the orbital separation of the satellite such that $v_{esc} \gtrsim
v(a)$. We need to estimate the tidal Love number, $k$, for Haumea in order to
evaluate the orbital evolution timescale. We infer from Haumea's density of
$\sim 3\rm{g/cm^3}$ that it must be mainly composed of rock. Using the
rigidity of basalt rock, $ \mu \sim 2 \times 10^{11}\rm{erg/cm^3}$
\citep{BA99}, we find $k=1.5/(1+\widetilde{\mu}) \sim 0.01$ where
$\widetilde{\mu}$ is the effective rigidity given by $\widetilde{\mu}=57
\mu/(8\pi\rho^2GR^2)$. Evaluating the tidal evolution timescale in equation
(\ref{5e31}) we find that the timescale for the semi-major axis to increase
such that $v(a) \sim v_{esc}$ is $\sim 6 (Q/100)(0.01/k) \rm{Myr}$. The
satellite therefore has sufficient time to undergo tidal evolution that
increases its semi-major axis such that $v_{esc} \gtrsim v(a)$ before it
suffers a destructive collision. Due to the long collision timescale, the
satellite will most likely undergo tidal evolution for $\sim 1$ Gyr before it
is broken apart. This yields a satellite separation from Haumea at the time of
the satellite break up of $\sim 17000\rm{km}$.
 
Haumea's spin angular momentum provides an upper limit on the mass of the
initial satellite that was later broken up into the family members. Assuming
no angular momentum was added to the system after the giant impact and that
Haumea was initially spinning close to break up, we find a maximum satellite
orbital angular momentum of $\sim 4 \times 10^{36}~\rm{gcm^2/s}$. For a
destructive satellite collision at an orbital separation of $\sim
17000~\rm{km}$ this yields a satellite mass of $\sim 2 \times
10^23~\rm{g}$. Our formation scenario, therefore, predicts that the total mass
of all the family members combined should not exceed $\sim 1/20$ of the mass
of Haumea or about 3 times the mass of the $R\sim 260~\rm{km}$ satellite
considered in the calculation for the family forming event above.

In summary, we propose that Haumea suffered a giant impact that leads to the
formation of a large, $\sim 260$km radius, satellite. Tidal evolution
increases the semi-major axis of the satellite such that $v_{esc} \gtrsim
v(a)$ in $\sim 10^7$ years. The satellite suffers a destructive collision with
an unbound KBO. This collision breaks the satellite into the different family
members and ejects them from the Haumea system. This results in a typical
velocity dispersion among the family members of $\sim 190$m/s. We propose that
Hi'iaka and Namaka are remnants of this collision that did not escape from the
Haumea system. The destructive satellite collision that leads
to the formation of the Haumea family has a collision timescale of several
tens of Gyrs which makes our formation scenario consistent with having one
collisional family for Haumea-sized objects in the Kuiper belt.

\subsection{Formation of Multiple Satellites and Ejection by Collisions with unbound KBOs} 

One can imagine that Haumea's initial giant impact did not generate just one
but several tightly bound satellites. The newly formed satellites undergo
tidal evolution that increases their orbital separation from Haumea. Once the
orbital separation is sufficiently large, the majority of the satellites
become gravitationally unbound from Haumea due to collisions with small,
unbound KBOs. In this case, the typical velocity dispersion of the family will
be of the order of the orbital velocity around Haumea before ejection which we
show is $\sim 120 \rm{m/s}$.

The rate at which a given satellite suffers collisions with unbound KBOs is
given by equation (\ref{5e32}). The satellite sizes and impactor size needed
to eject a satellite by collion however differ from that required for
satellite destruction in the previous section. In a given collision,
the velocity change of the satellite is given by the conservation of linear
momentum, $\Delta v(a) r_s^3= \chi r^3 v_{disp}$. The coefficient $\chi$
accounts for the final momentum of the impactor. If the unbound KBO is
perfectly reflected $\chi=2$. Momentum loss from an impact crater can lead to
$\chi>2$ where the exact value of $\chi$ depends on the properties of the
colliding bodies \citep{MNZ94}. Since we are primarily concerned with deriving
an order of magnitude estimate for the impactor size we will adopt $\chi=1$ for
the rest of this paper. A satellite of Haumea can be ejected by a collision
with an unbound KBO if it suffers a velocity change $\Delta v(a) \sim
v(a)$. Therefore, in order to be ejected, a satellite needs to collide with a
KBO that typically has a radius of $r \sim r_s (v(a)/v_{disp})^{1/3}$. 

Substituting this expression for $r$ into equation (\ref{5e32}) we have find
that the ejection timescale for Haumea's satellites by a collision with an
unbound KBO is given by
\begin{equation}\label{5e35}
\tau_{coll}=\mathcal{R}_{coll}^{-1} \sim
\frac{\rho r_s}{\Sigma}
\left(\frac{v(a)}{v_{disp}}\right) \Omega^{-1}.
\end{equation}
For the ejection of Haumea's satellites by collisions with unbound KBOs to be
the typical outcome we need $\tau_{coll} \sim \tau_{tidal}$ since otherwise
most of the satellites should have remained bound to Haumea which is
contradicted by observations of the Haumea family. Equating the tidal
evolution and ejection timescales allows us to derive the typical velocity
with which the family members left the Haumea system. Equating equations
(\ref{5e31}) and (\ref{5e35}) and solving for $v(a)$ we have
\begin{equation}\label{5e36}
v(a) \sim v_B^{13/14} v_{disp}^{1/14} \left[ \left(\frac{2}{39}\right)
\left(\frac{Q}{k}\right) \left(\frac{M}{m_s}\right)
\left(\frac{\Sigma}{\rho r_s}\right)
\left(\frac{\Omega}{\Omega_B}\right)
\right]^{1/14} \sim 120\rm{m/s}
\end{equation}
where we used $Q/k \sim 10^4$ and $m_s \sim 4.2 \times 10^{21}$g (i.e. $M/m_s
\sim 10^3$) to estimate $v(a)$. Evaluating equation (\ref{5e36}) for the
various masses of the family members we find that $v(a)$ ranges from $98
\rm{m/s}$ to $178 \rm{m/s}$ with a typical value of $\sim 120
\rm{m/s}$. Therefore, Haumea's satellites will be ejected from the Haumea
system by collisions with unbound KBOs once their orbital velocity around
Haumea is $\sim 120$m/s. This will also be roughly, the expected velocity
dispersion between family members which is in good agreement with the
observations \citep{BB07,RB07}.

Evaluating both the collisional and tidal evolution timescale using $v(a)\sim
120m/s$ we find from equations (\ref{5e35}) and (\ref{5e31}) that the typical
ejection timescale is $\sim 60 \rm{Gyr}$. The ejection
timescales for the various masses of the family members are all tens of Gyrs
and therefore exceed the age of the solar system. Our calculation here can
only estimate the ejection timescale to an oder of magnitude, it might
therfore be that the actual ejection timescale is somewhat shorter than
estimate here. In addition, we might have underestimated the number of unbound
KBOs that can lead to the ejection of family members since we extrapolated the
surface density of 100km sized bodies to smaller sizes assuming a power-law
index $q$ of 4 whereas the actual power-law index might be a little larger
than this. Since this second formation scenario for the Haumea family involves
the ejection of all the family members separately the ejection process cannot
be a rare event. The ejection timescales therefore need to be shorter than the
age of the solar system for this formation scenario to be feasible. This
however raises a different problem. If the ejection timescales are indeed
shorter than the age of the solar system, then most of the binaries in the
Kuiper belt should have been broken apart by the same process. This is in
contradiction with the observations and we therefore conclude that formation
of the Haumea family by a destructive collision of a single satellite is the
preferred scenario.

In addition to problems discussed above, this second scenario faces yet
another challenge. If the initial giant impact of Haumea produced several
satellites then satellite-satellite interactions need to be taken into
account. The timescale for satellite ejection due to satellite-satellite
interactions is $\sim (M/m_s)^2 a/v(a)$ \citep{GLS04}. This timescale is very
short (i.e. $\sim 4 \times 10^3 \rm{years}$ for $M/m_s \sim 10^3$ and a 10 day
satellite orbit). Initially, however, $v(a) > v_{esc}$ which implies that the
satellites tend to collide with each other rather than eject each other from
the system. Satellite-satellite collisions may either lead to accretion or
break up. In either case it is questionable whether several satellites can
survive and tidally evolve outwards such that they could be ejected by
collisions with unbound KBOs. Satellite-satellite interactions are therfore
yet another reason to favor our first scenario in which a single satellite is
created and broken apart.

As an alternative to ejecting the satellites by collisions with unbound KBOs,
one can imagine that the satellites could have been removed from the Haumea
system by gravitational scattering of passing KBOs. However, in the high
velocity regime discussed here ($v_{disp}>v_{esc}$), the rate of satellite
ejection due the gravitational scattering is much less than that due to direct
impacts of unbound KBOs onto the satellites. See \citet{CS08} for comparison
of collisional and gravitational evolution of binaries.

\section{HAUMEA'S INITIAL GIANT IMPACT}\label{5s3}
\citet{BB07} estimate that Haumea's radius before its giant impact was $\sim
830$km and that the impactor was $\sim 500$km in radius. The timescale for
such an impact to occur in todays Kuiper belt can be found from equation
(\ref{5e32}) which yields a collision timescale of $\sim 8 \times 10^{12}$
years when evaluated for $R=830$km and $r=500$km. Such a collision is
therefore extremely unlikely but needed if one wants to form and eject the
family directly form the giant impact. \citet{Lev08} propose a giant impact
scenario for Haumea that circumvents this low probability by requiring a
collision between two scattered disk objects assuming that the scattered disk
was a 100 times more massive than it is today.  In our formation scenario
Haumea can suffer its giant impact before the Kuiper belt was dynamically
excited which shortens the collision timescale significantly. The timescale
for Haumea's giant impact in the sub-Hill velocity regime is
\citep{GLS04,SR07}
\begin{equation}\label{5e41}
\tau_{coll}\sim \frac{\rho r^3}{\Sigma\Omega R^{2}} \alpha^{3/2} \sim
8 \times 10^6\rm{years}
\end{equation} 
where $\alpha=R/R_H \sim 10^{-4}$ and $R_H$ is Haumea's Hill
radius. Therefore, allowing Haumea's giant impact to occur while the Kuiper
Belt was still dynamically cold decreases the giant impact timescale by 6
orders of magnitude, even without enhancing the mass surface density in the
Kuiper Belt above its estimated current value. We therefore propose that
Haumea's initial giant impact occurred while the velocity dispersion of large
KBOs was still in the sub-Hill regime. This is supported by the ubiquity of
small, collisionally formed satellites around KBOs, which have radii as large
as 1000km \citep{BDB06,BS07} and the Pluto-Charon system \citep{WSM06} which
strongly suggests that sub-Hill KBO velocities prevailed during satellite
formation and that collisional satellite formation was common, especially
around the largest KBOs. The satellite, which we propose forms in Haumea's
giant impact, is initially tightly bound to Haumea and the long tidal
evolution timescale ensures that the Haumea-satellite system remains intact
until after the dynamical excitation of the Kuiper belt. The family members
are created and ejected from the Haumea system only after the dynamical
excitation of the Kuiper belt which ensures the dynamical coherence of the
family members. This scenario, therefore, does not face the potential
challenge of removing 99\% of the mass in the scattered disk without
destroying the dynamical coherence of the family.

\section{DISCUSSION AND CONCLUSION}\label{5s4}
We propose a new formation scenario for the Haumea family. In our scenario
Haumea's giant impact forms a single moon that tidally evolves outward until
it suffers a destructive collision from which the family is created. The
advantage of this scenarios is that it naturally gives rise to a lower
velocity dispersion among the family members than expected from direct
ejection from Haumea's surface. We show that this formation scenarios yields a
velocity dispersion of $\sim 190 \rm{m/s}$ among the family members. This is
in good agreement with the measured dispersion $\sim 140$m/s in semi-major
axis, eccentricity, and inclination of the family members \citep{BB07,RB07}
which is a lower limit of the actual velocity dispersion since the orbital
angles were chosen to minimize the velocity dispersion of the family
\citep{RB07}. Our formation scenario yields about one collisional family
for Haumea-sized objects in the Kuiper Belt. Ejecting the family members from
Haumea's orbit has the additional advantage that it is easy to reconcile with
Haumea's initial giant impact. The family must have been ejected from Haumea
after the Kuiper belt was dynamically excited in order to preserve the
dynamical coherence of the family. If the family members are dispersed
fragments of the giant impact itself then, the giant impact must have occurred
after the Kuiper belt was dynamically excited. Such a giant impact occurs with
a probably of less than $10^{-3}$ over the age of the solar system and is
therfore extremely unlikely in todays Kuiper belt. \citet{Lev08} suggest that
Haumea's giant impact could be the result of collision between two scattered
disk objects during a phase when the scattered disk was a 100 times more
massive than it is today. In our scenario, Haumea's giant impact can occur
before the dynamical excitation of the Kuiper Belt since the giant impact and
the ejection of the family are two different events separated in time by at
least $\sim 10^7$years. The timescale for Haumea's giant impact in the
sub-Hill velocity regime is $\sim 8 \times 10^6$ years. Observations show that
the majority of the largest KBOs have small, collisionally formed satellites
\citep{BDB06,BS07}. Giant impacts that lead to satellite formation around
large Kuiper Belt objects were therefore common in the history of the Kuiper
Belt and we propose that Haumea's initial giant impact was one of them. Our
formation scenario is also in agreement with results from simulations of giant
impacts since it only requires the formation of a satellite and not the
simultaneous formation of satellites and direct ejection of fragments in a
single collision \citep{MR97,BA99,AA04,C04,C05} which is required in the
original formation scenario proposed by \citet{BB07}.

In addition to the family members discussed above, Haumea has also two
satellites. Hi'iaka the larger outer satellite ($M/m_s \sim 200$) has a
semi-major axis of 49500km and a free eccentricity of 0.07 \citep{BB05,RB09}.
Namaka, the smaller ($M/m_s \sim 2000$) inner satellite, has a semi-major axis
of 25700km, a free eccentricity of 0.21 and its inclination with respect to
Hi'iaka is $13^{\circ}$ \citep{RB09}. Hi'iaka and Namaka display, just like
all other family members, strong water ice absorption features in their
infrared spectra \citep{BBS06,FB09}. Since this spectral signature seems to be
only present among the family members it seems unlikely that Hi'iaka and
Namaka were captured; instead they most likely formed together with the other
family members. It is unlikely that Hi'iaka and Namaka evolved to their
current separation by tides, since the tidal evolution timescales are
excessively long. From equation (\ref{5e31}) we have for Hi'iaka $\tau_{tidal}
\sim 4 \times 10^{12}$ years and for Namaka $\tau_{tidal} \sim 6 \times
10^{11}$ years where we used again $Q \sim 100$ and $k\sim 0.01$. Both
timescales exceed the age of the solar system by more than two orders of
magnitude. We suggest that Hi'iaka and Namaka were produced in the same
satellite-break-up that created the other family members, only that in their
case the impulse was not sufficient to escape Haumea but instead it increased
their semi-major axis by a factor of $\sim 2$ to their current
separation. Such a collision will however also raise the eccentricity to order
unity. Is is possible that the satellites, especially Hi'iaka, formed by
re-accumulation of collisional fragments of the satellite break up. Such a
re-accumulation scenario typically leads to more circular satellite orbits. We
also note that Hi'iaka's free eccentricity of 0.07 is consistent with
dynamical excitations by passing KBOs \citep{CS08}. Namaka, which is ten times
less massive than Hi'iaka, could be a single collisional fragment of the
satellite break up, hence its large free eccentricity of 0.21. We therefore
find that our formation scenario for Haumea's family can account for the large
semi-major axis and modest eccentricities of Hi'iaka and Namaka. The $13^{
\circ}$ mutual inclination between the two moons remains somewhat of a puzzle,
since it is surprisingly high if the moons formed in a disk and tidally
evolved outward and it is surprisingly low if the moons formed from fragments
of a disruptive satellite collision as suggested in this paper.

\acknowledgments We would like to thank Darin Ragozzine, Mike Brown, and
Yanqin Wu for valuable discussions. R. S. is a Packard Fellow. This research
was partially supported by the ERC.

\end{document}